\documentclass[sigconf]{acmart}

\usepackage{listings}
\usepackage{multirow}
\usepackage{pifont}
\usepackage{balance}

\AtBeginDocument{%
  \providecommand\BibTeX{{%
    \normalfont B\kern-0.5em{\scshape i\kern-0.25em b}\kern-0.8em\TeX}}}

\copyrightyear{2024}
\acmYear{2024}
\setcopyright{acmlicensed}
\acmConference[WWW '24 Companion]{Companion Proceedings of the ACM Web Conference 2024}{May 13--17, 2024}{Singapore, Singapore}
\acmBooktitle{Companion Proceedings of the ACM Web Conference 2024 (WWW '24 Companion), May 13--17, 2024, Singapore, Singapore}
\acmDOI{10.1145/3589335.3651463}
\acmISBN{979-8-4007-0172-6/24/05}

\begin{document}

\title{A Study of Vulnerability Repair in JavaScript Programs with Large Language Models}

\author{Tan Khang Le}
\orcid{0009-0002-1703-4066}
\affiliation{%
  \institution{Simon Fraser University}
  \city{Burnaby}
  \state{British Columbia}
  \country{Canada}
  \postcode{V5A 1S6}
}
\email{khang_le@sfu.ca}

\author{Saba Alimadadi}
\orcid{0000-0002-5667-152X}
\affiliation{%
  \institution{Simon Fraser University}
  \city{Burnaby}
  \state{British Columbia}
  \country{Canada}
  \postcode{V5A 1S6}
}
\email{saba@sfu.ca}

\author{Steven Y. Ko}
\orcid{0000-0003-3771-0156}
\affiliation{%
  \institution{Simon Fraser University}
  \city{Burnaby}
  \state{British Columbia}
  \country{Canada}
  \postcode{V5A 1S6}
}
\email{steveyko@sfu.ca}

\begin{abstract}
In recent years, JavaScript has become the most widely used programming language, especially in web development. However, writing secure JavaScript code is not trivial,  and programmers often make mistakes that lead to security vulnerabilities in web applications. Large Language Models (LLMs) have demonstrated substantial advancements across multiple domains, and their evolving capabilities indicate their potential for automatic code generation based on a required specification, including automatic bug fixing. In this study, we explore the accuracy of LLMs, namely ChatGPT and Bard, in finding and fixing security vulnerabilities in JavaScript programs. We also investigate the impact of context in a prompt on directing LLMs to produce a correct patch of vulnerable JavaScript code. Our experiments on real-world software vulnerabilities show that while LLMs are promising in automatic program repair of JavaScript code, achieving a correct bug fix often requires an appropriate amount of context in the prompt.
\end{abstract}



\maketitle

\section{Introduction}
Despite the prevalence and popularity of JavaScript programs, understanding and analyzing them is challenging due to their heterogeneous, dynamic, and asynchronous nature. As a result, developers are prone to making mistakes and exposing many JavaScript programs to security vulnerabilities \cite{crockford2008javascript}. Developers commonly use static analysis and fuzzing techniques to mitigate such bugs. However, this process can be tedious in terms of understanding and identifying the vulnerabilities, and subsequently modifying the code to repair the bugs.

With the rapid development of artificial intelligence, Large Language Models (LLMs) are increasingly trained on large codebases with the goal of automatic code generation based on specifications from user inputs \cite{austin2021program, chowdhery2023palm, nijkamp2022codegen, xu2022systematic}. This empowers LLMs to generate code in different ways, given some context such as the developer's intention expressed in code comments. Although LLMs may occasionally produce code with security bugs \cite{siddiq2022empirical, sandoval2023lost}, when coupled with suitable security-aware tooling during code generation, LLMs have the potential to enhance a software developer's productivity \cite{pearce2022asleep}, thereby reducing the risks of introducing new security bugs. As such, the research community has been actively investigating the effectiveness of LLMs in finding and fixing code vulnerabilities \cite{pearce2023examining, he2023large, tol2023zeroleak, wu2023effective, ahmad2023fixing}. Most of these studies, however, have focused on programming languages such as C/C++ and Verilog. As such, we currently do not have much insight into the role of LLMs in repairing security vulnerabilities in a dynamic language such as JavaScript.

In this paper, we study the utilization of black-box, ``off-the-shelf" LLMs, namely ChatGPT and Bard, in the automatic program repair of JavaScript code. Furthermore, we investigate the effect of context (or \textit{cues}) in a prompt on LLMs' ability to generate accurate security patches. We aim to address the following research questions:
\begin{itemize}
    \item \textbf{RQ1}: How accurate are LLMs in finding and fixing vulnerabilities in JavaScript programs?
    \item \textbf{RQ2}: How does the amount of context in a prompt impact the effectiveness of LLMs in producing a correct patch of vulnerable JavaScript code?
\end{itemize}

To answer these questions, we conduct a study on over 20 of the most common software vulnerabilities. We compare three distinct prompt templates with varying degrees of contextual \textit{cues}. These templates serve to guide LLMs in repairing vulnerable JavaScript code. In total, the study involves 60 prompts for two LLMs, covering the 20 identified vulnerabilities across the three proposed prompt templates\footnote{Our repair prompts and testing results are publicly available at: \url{https://doi.org/10.5281/zenodo.10783763}}. The experimental results show that ChatGPT and Bard, on average, accurately generate patches for 71.66\% and 68.33\% of cases, respectively (RQ1). Furthermore, we find that the more context provided in a prompt, the better ChatGPT and Bard perform in producing a correct patch, with an improvement of up to 55\% in accuracy (RQ2).

\section{Related Work}
In this section, we review the related work on code security and vulnerability as well as program repair with large language models.

\subsection{Code Security and Vulnerability}
Due to the critical importance of safeguarding systems against potential threats and breaches, there has been much research into code security and vulnerability, providing insights into the nature of security bugs and weaknesses. Specifically, as system errors and vulnerabilities continue to grow over time, automated program repair \cite{le2021automatic} emerges as a research field that focuses on a class of techniques for producing source code-level patches for such bugs. A classical approach for automated program repair is to turn the program repair problem into a search problem. For example, \citet{le2011genprog} used genetic programming to search for a program variant that addresses a bug in the given program without changing the required functionality or producing new errors. An alternative approach demonstrated by \citet{nguyen2013semfix} is to use semantic analysis for the automatic construction of patches.

The development community also actively engages in studying and documenting vulnerabilities in software and hardware, such as the Common Weakness Enumeration (CWE). CWE functions as a categorization system designed for both hardware and software vulnerabilities, consisting of over 600 categories, namely buffer overflows, race conditions, and cross-site scripting. Every year, MITRE publishes a list of the top 25 most dangerous software weaknesses (CWE Top 25) \cite{mitre25}, which serves as a benchmark for evaluating many automated program repair tools.

\subsection{Program Repair with Large Language Models (LLMs)}
With the rise of LLMs in recent years, the field of automated program repair has seen many major advancements. Much research has shown the potential of LLMs in automated bug fixing for both software and hardware programs. Particularly, \citet{wu2023effective} evaluated five different LLMs and four automatic program repair tools on two Java vulnerability benchmarks. With regard to hardware vulnerabilities, \citet{ahmad2023fixing} attempted to repair hardware bugs in Verilog programs automatically using LLMs. From the security perspective, the study by \citet{pearce2023examining} on zero-shot vulnerability repair demonstrated promising results of using LLMs for finding and fixing security bugs in both software and hardware programs, written in Python, C, and Verilog. Furthermore, recent research has leveraged LLMs to produce highly secured code through controlled code generation \cite{he2023large} or to detect and patch side-channel vulnerabilities in software \cite{tol2023zeroleak}.

\section{Methodology}
In our study, given an input \textit{prompt} that instructs LLMs to find and fix a \textit{vulnerability} within a provided JavaScript code snippet, we investigate the accuracy of the responses generated by these LLMs. An example of an input prompt and the corresponding response returned by ChatGPT are shown in Table \ref{table:prompt-and-response-example}.

\begin{table}
    \caption{An example of an input prompt and the response returned by LLM.}
    \label{table:prompt-and-response-example}
    \begin{center}
        \begin{tabular}{ p{3.9cm} p{3.9cm} } 
            \toprule
            \multicolumn{1}{c}{\textbf{Prompt}} & \multicolumn{1}{c}{\textbf{Response}} \\
            \midrule
            \begin{lstlisting}
Fix the bug in the following code:

```js
let idSequence = new Int16Array(3);

idSequence[0] = 123;
idSequence[1] = 234;
idSequence[2] = 345;
idSequence[3] = 456; 
    // No effect

console.log(idSequence[3]) 
    // undefined
```
            \end{lstlisting} &
            \begin{lstlisting}
Here is the corrected code:

```js
let idSequence = new Int16Array(4); 
    // Fix the length to 4

idSequence[0] = 123;
idSequence[1] = 234;
idSequence[2] = 345;
idSequence[3] = 456;

console.log(idSequence[3]);
    // Accessing the valid index 3
```
            \end{lstlisting} \\
            \bottomrule
        \end{tabular}
    \end{center}
\end{table}
 
In the given example, we ask LLMs to fix a bug, namely an ``Out-of-bound Write" error, in the given JavaScript code. Particularly, the issue in the code is that an \texttt{Int16Array} object is created with a length of \texttt{3}, but there is a write operation to the element at index \texttt{3}, which will take no effect because the valid indices in this array are \texttt{0}, \texttt{1}, and \texttt{2}. In this scenario, ChatGPT successfully identifies and fixes the bug by increasing the array length to \texttt{4}, effectively validating element access at index \texttt{3}. We note that various approaches may exist for repairing vulnerable code. However, as long as LLMs can produce a functional and secured patch, we consider it a success as our focus lies in determining the capability of LLMs to identify a solution rather than exhaustively exploring all potential fixes.

\subsection{Prompt Design}
To investigate the role of different levels of context in the prompts on the generated repair, we design three different prompt templates based on the degree of provided context. First, a context-free (\texttt{c.f.}) prompt comprises solely a repair instruction and a vulnerable JavaScript code. There are no hints or additional information regarding the anticipated vulnerability. Second, a context-sensitive (\texttt{c.s.}) prompt is similar to a context-free prompt but with the inclusion of the name of the expected vulnerability, such as ``Out-of-bound Write" and ``Integer Overflow." Last, a context-rich (\texttt{c.r.}) prompt encloses comments along with the vulnerable JavaScript code, providing an exhaustive explanation of the vulnerability within the code as well as how an attacker could exploit it. While in an ideal scenario, a developer with the knowledge of a vulnerability context should be able to repair it directly without relying on LLMs, explaining the vulnerability to LLMs as in \texttt{c.r.} prompts might be beneficial for seeking suggestions or exploring different perspectives on potential fixes. This collaborative approach could enhance a developer's productivity in vulnerability repair. A sample prompt for each template is shown in Table \ref{table:sample-prompts}.

\begin{table}
    \centering
    \caption{Samples of the proposed prompt templates.}
    \label{table:sample-prompts}
    \begin{center}
        \begin{tabular}{ p{1.2cm} p{6.4cm} } 
            \toprule
            \multicolumn{1}{c}{\textbf{Template}} & \multicolumn{1}{c}{\textbf{Sample}} \\
            \midrule
            \begin{flushleft}
                context-free (\texttt{c.f.})
            \end{flushleft} & \begin{lstlisting}
Fix the bug in the following code:

```js
const urlParams = new URLSearchParams(window.location.search);
const username = urlParams.get('username');
const trimmedName = username.trim();
```
            \end{lstlisting} \\\hline
            \begin{flushleft}
                context-sensitive (\texttt{c.s.})
            \end{flushleft} & \begin{lstlisting}
Fix the "NULL Pointer" vulnerability in the following code:

```js
const urlParams = new URLSearchParams(window.location.search);
const username = urlParams.get('username');
const trimmedName = username.trim();
```
            \end{lstlisting} \\\hline
            \begin{flushleft}
                context-rich (\texttt{c.r.}) 
            \end{flushleft} & \begin{lstlisting}
Fix the bug in the following code:

```js
// Get the query parameters from the current URL
const urlParams = new URLSearchParams(window.location.search);
// Get the username value
const username = urlParams.get('username');
// Trim any leading and ending white spaces in username
const trimmedName = username.trim();

// Users can control the query parameters so an attacker may set username value to undefined or null
// causing the program to crash when it attempts to call the trim method
```
            \end{lstlisting} \\
            \bottomrule
        \end{tabular}
    \end{center}
\end{table}

\subsection{Vulnerability Selection}
\begin{table}
    \caption{Relevant vulnerabilities selected from 2023 CWE Top 25 List.}
    \label{table:selected-vulnerabilities}
    \begin{center}
        \begin{tabular}{ p{1.3cm} p{6.4cm} } 
            \toprule
            \multicolumn{1}{c}{\textbf{ID}} & \multicolumn{1}{c}{\textbf{Description}} \\
            \midrule
            CWE-20 & Improper Input Validation \\\hline
            CWE-22 & Improper Limitation of a Pathname to a Restricted Directory (`Path Traversal') \\\hline
            CWE-77 & Improper Neutralization of Special Elements used in a Command (`Command Injection') \\\hline
            CWE-78 & Improper Neutralization of Special Elements used in an OS Command (`OS Command Injection') \\\hline
            CWE-79 & Improper Neutralization of Input During Web Page Generation (`Cross-site Scripting') \\\hline
            CWE-89 & Improper Neutralization of Special Elements used in an SQL Command (`SQL Injection') \\\hline
            CWE-94 & Improper Control of Generation of Code (`Code Injection') \\\hline
            CWE-125 & Out-of-bounds Read \\\hline
            CWE-190 & Integer Overflow or Wraparound \\\hline
            CWE-269 & Improper Privilege Management \\\hline
            CWE-276 & Incorrect Default Permissions \\\hline
            CWE-287 & Improper Authentication \\\hline
            CWE-306 & Missing Authentication for Critical Function \\\hline
            CWE-434 & Unrestricted Upload of File with Dangerous Type \\\hline
            CWE-476 & NULL Pointer Dereference \\\hline
            CWE-502 & Deserialization of Untrusted Data \\\hline
            CWE-787 & Out-of-bounds Write \\\hline
            CWE-798 & Use of Hard-coded Credentials \\\hline
            CWE-862 & Missing Authorization \\\hline
            CWE-863 & Incorrect Authorization \\
            \bottomrule
        \end{tabular}
    \end{center}
\end{table}

To ensure that our study is practical and relevant to the real world, we leverage the latest 2023 CWE Top 25 List \cite{mitre25}. However, not all vulnerabilities listed in the top 25 list are related to JavaScript, a few of them are specific to other programming languages. For example, ``CWE-416: Use After Free" is only applicable to C and C++ as described in MITRE's documentation. Consequently, we carefully identify and select 20 out of the 25 vulnerabilities that are most relevant to JavaScript. The complete list of the identified 20 vulnerabilities is presented in Table \ref{table:selected-vulnerabilities}.

\section{Experiment and Evaluation}
In our study, we conduct a systematic experiment on two publicly available LLMs, namely ChatGPT and Bard.

\subsection{Experiment Details}
Based on the proposed three prompt templates and the identified 20 vulnerabilities, we formulate a total of 60 prompts. Specifically, each of the 20 vulnerabilities is replicated across the three prompt templates with varying degrees of contextual \textit{cues}, ranging from no additional context to comprehensive detail.

In our experiment, we feed our prompts to the LLMs and subsequently evaluate the correctness of their responses in repairing the anticipated vulnerability within the given JavaScript code snippet. Additionally, as LLMs can be heavily biased in a continuous dialogue \cite{zhao2023survey}, we create a new and separate conversation for each prompt to avoid such biases and ensure an impartial evaluation.

\subsection{Results and Discussion}
\noindent \textit{RQ1: How accurate are LLMs in finding and fixing vulnerabilities in JavaScript programs?}

Table \ref{table:performance-apr-llms} presents the performance results of ChatGPT and Bard on repairing various vulnerabilities, with different levels of context, in JavaScript programs. Particularly, ChatGPT correctly finds and fixes 43 out of 60 cases, achieving an accuracy of 71.66\%. On the other hand, Bard has a slightly lower accuracy of 68.33\%, with 41 accurate repairs out of 60.

\noindent \textit{RQ2: How does the amount of context in a prompt impact the effectiveness of LLMs in producing a correct patch of vulnerable JavaScript code?}

The results in Table \ref{table:performance-apr-llms} also demonstrate that the provided context in a prompt has a positive impact on LLMs' capability to find and fix vulnerabilities. When there is no additional context as in \texttt{c.f.} prompts, ChatGPT and Bard perform poorly in repairing security bugs, each achieving an accuracy of 40\%. However, when compared with \texttt{c.f.} prompts, ChatGPT shows improved performance on \texttt{c.s.} and \texttt{c.r.} prompts, showcasing an improved accuracy of 80\% and 95\%, respectively. Similarly, Bard experiences a better accuracy of 80\% and 85\% when tested with \texttt{c.s.} and \texttt{c.r.} prompts, respectively.

\begin{table}
    \caption{Performance results of large language models on finding and fixing vulnerabilities in JavaScript code.}
    \label{table:performance-apr-llms}
    \begin{center}
        \begin{tabular}{ l l l l l l l } 
            \toprule
            \multirow{2}{*}[-2pt]{\parbox{1.5cm}{Vulnerability ID}} & \multicolumn{3}{c}{\textbf{ChatGPT}} & \multicolumn{3}{c}{\textbf{Bard}} \\
                                            \cmidrule(lr){2-4} \cmidrule(lr){5-7}
                                            & \multicolumn{1}{c}{\texttt{c.f.}} & \multicolumn{1}{c}{\texttt{c.s.}} & \multicolumn{1}{c}{\texttt{c.r.}} & \multicolumn{1}{c}{\texttt{c.f.}} & \multicolumn{1} {c}{\texttt{c.s.}} & \multicolumn{1}{c}{\texttt{c.r.}} \\
            \midrule
            \textbf{CWE-20} & \color{red}\ding{55} & \color{green}\ding{51} & \color{green}\ding{51} & \color{red}\ding{55} & \color{green}\ding{51} & \color{green}\ding{51} \\
            \textbf{CWE-22} & \color{red}\ding{55} & \color{green}\ding{51} & \color{green}\ding{51} & \color{red}\ding{55} & \color{green}\ding{51} & \color{green}\ding{51} \\
            \textbf{CWE-77} & \color{green}\ding{51} & \color{green}\ding{51} & \color{green}\ding{51} & \color{green}\ding{51} & \color{green}\ding{51} & \color{green}\ding{51} \\
            \textbf{CWE-78} & \color{green}\ding{51} & \color{green}\ding{51} & \color{green}\ding{51} & \color{red}\ding{55} & \color{green}\ding{51} & \color{green}\ding{51} \\
            \textbf{CWE-79} & \color{green}\ding{51} & \color{green}\ding{51} & \color{green}\ding{51} & \color{green}\ding{51} & \color{green}\ding{51} & \color{green}\ding{51} \\
            \textbf{CWE-89} & \color{green}\ding{51} & \color{green}\ding{51} & \color{green}\ding{51} & \color{green}\ding{51} & \color{green}\ding{51} & \color{green}\ding{51} \\
            \textbf{CWE-94} & \color{green}\ding{51} & \color{green}\ding{51} & \color{green}\ding{51} & \color{green}\ding{51} & \color{green}\ding{51} & \color{green}\ding{51} \\
            \textbf{CWE-125} & \color{red}\ding{55} & \color{green}\ding{51} & \color{green}\ding{51} & \color{red}\ding{55} & \color{green}\ding{51} & \color{green}\ding{51} \\
            \textbf{CWE-190} & \color{red}\ding{55} & \color{red}\ding{55} & \color{green}\ding{51} & \color{red}\ding{55} & \color{red}\ding{55} & \color{red}\ding{55} \\
            \textbf{CWE-269} & \color{red}\ding{55} & \color{red}\ding{55} & \color{red}\ding{55} & \color{red}\ding{55} & \color{green}\ding{51} & \color{green}\ding{51} \\
            \textbf{CWE-276} & \color{green}\ding{51} & \color{green}\ding{51} & \color{green}\ding{51} & \color{green}\ding{51} & \color{green}\ding{51} & \color{green}\ding{51} \\
            \textbf{CWE-287} & \color{red}\ding{55} & \color{red}\ding{55} & \color{green}\ding{51} & \color{red}\ding{55} & \color{green}\ding{51} & \color{green}\ding{51} \\
            \textbf{CWE-306} & \color{red}\ding{55} & \color{red}\ding{55} & \color{green}\ding{51} & \color{red}\ding{55} & \color{green}\ding{51} & \color{green}\ding{51} \\
            \textbf{CWE-434} & \color{red}\ding{55} & \color{green}\ding{51} & \color{green}\ding{51} & \color{red}\ding{55} & \color{green}\ding{51} & \color{green}\ding{51} \\
            \textbf{CWE-476} & \color{green}\ding{51} & \color{green}\ding{51} & \color{green}\ding{51} & \color{green}\ding{51} & \color{red}\ding{55} & \color{red}\ding{55} \\
            \textbf{CWE-502} & \color{red}\ding{55} & \color{green}\ding{51} & \color{green}\ding{51} & \color{green}\ding{51} & \color{green}\ding{51} & \color{green}\ding{51} \\
            \textbf{CWE-787} & \color{green}\ding{51} & \color{green}\ding{51} & \color{green}\ding{51} & \color{green}\ding{51} & \color{green}\ding{51} & \color{green}\ding{51} \\
            \textbf{CWE-798} & \color{red}\ding{55} & \color{green}\ding{51} & \color{green}\ding{51} & \color{red}\ding{55} & \color{red}\ding{55} & \color{red}\ding{55} \\
            \textbf{CWE-862} & \color{red}\ding{55} & \color{green}\ding{51} & \color{green}\ding{51} & \color{red}\ding{55} & \color{red}\ding{55} & \color{green}\ding{51} \\
            \textbf{CWE-863} & \color{red}\ding{55} & \color{green}\ding{51} & \color{green}\ding{51} & \color{red}\ding{55} & \color{green}\ding{51} & \color{green}\ding{51} \\
            \midrule
            Accuracy & \textbf{8/20} & \textbf{16/20} & \textbf{19/20} & \textbf{8/20} & \textbf{16/20} & \textbf{17/20} \\
            \bottomrule
        \end{tabular}
    \end{center}
\end{table}

\subsection{Threats to Validity}
The validity of the evaluations drawn from our experimental results is subject to a few threats. First, we use the public version of both ChatGPT \cite{chatgpt} and Bard \cite{bard} in our testing (accessed in November 2023), which can change and evolve over time. Second, the correctness of the repaired JavaScript code produced by LLMs is verified manually and, thus, subject to human biases. Last, we acknowledge that more studies are required to draw more accurate conclusions.

\section{Conclusion}
In this study, we have identified 20 common vulnerabilities from the CWE Top 25 List that are most relevant to JavaScript and proposed three prompt templates with varying degrees of context. Based on these vulnerabilities and templates, we have formulated a total of 60 repair prompts for our study with LLMs, namely ChatGPT and Bard. Our experimental results show that ChatGPT excels with a promising accuracy of 71.66\%, while Bard closely follows with an accuracy of 68.33\%, in the automatic program repair task of vulnerable JavaScript code. Furthermore, our findings indicate that increased contextual information in a repair prompt positively influences the performance of LLMs in finding and fixing vulnerabilities, leading to a significant boost in accuracy of up to 55\%.


\bibliographystyle{ACM-Reference-Format}
\balance
\bibliography{sample-base}

\end{document}